\documentstyle[prl,aps,preprint,tighten,floats,epsfig,psfig]{revtex}

\newcommand{\beq}{\begin{equation}}
\newcommand{\eeq}{\end{equation}}
\newcommand{\bea}{\begin{eqnarray}}
\newcommand{\eea}{\end{eqnarray}}

\begin{document}
\draft
\date{\today}
\preprint{\vbox{\baselineskip=13pt
\rightline{LPT Orsay-99/16}
}}

\title{Asymptotic behaviour of the gluon propagator from lattice QCD}
\author{ D. Becirevic$^a$, Ph. Boucaud$^b$,  J.P. Leroy$^b$, 
J. Micheli$^b$, \\  O. P\`ene$^b$, J. Rodr\'\i guez--Quintero$^b$  
and C. Roiesnel$^c$ } \par \maketitle
\begin{center}
$^a$INFN, Sezione di Roma, P.le Aldo Moro 2, I-00185 Rome, Italy \\
$^b$Laboratoire de Physique Th\'eorique \footnote{Unit\'e Mixte de Recherche
 - UMR 8627}\\
{Universit\'e de Paris XI, B\^atiment 211, 91405 Orsay Cedex,
France}\\$^c$ Centre de Physique Th\'eorique\footnote{
Unit\'e Mixte de Recherche C7644 du Centre National de 
la Recherche Scientifique\\ 
\\e-mail: Philippe.Boucaud@th.u-psud.fr, roiesnel@cpht.polytechnique.fr
}de l'Ecole Polytechnique\\
91128 Palaiseau Cedex, France 

\end{center}

\begin{abstract}

We study the flavorless gluon propagator in the Landau gauge
from high statistics lattice calculations. 
Hypercubic artifacts are efficiently eliminated by taking the
$\sum p_\mu^4 \to 0$ limit. The propagator  is fitted to the
three-loops perturbative formula in an  energy window
ranging form $\sim$ 2.5 GeV up to $\sim$ 5.5 GeV.
$\alpha_s$ is extracted from the best fit in a continuous set of
 renormalisation
 schemes. The fits are very good, with a $\chi^2$ per
d.o.f smaller than 1. We propose a more stringent test of asymptotic
scaling based on scheme independence of the resulting 
$\Lambda_{\overline {\rm MS}}$. This method  shows 
 that asymptotic scaling at three loops is not reached 
by the gluon propagator although we use rather large energies. 
We are only able to obtain an
effective flavorless three-loops estimate 
$\Lambda_{\overline {\rm MS}}^{(3)}= 353 \pm 2 ^{+25}_{-10}\,$ MeV.
 We argue that the real asymptotic value for $\Lambda_{\overline {\rm
 MS}}$ should plausibly be smaller.
 
\end{abstract}
\begin{flushright} LPT Orsay-99/16\\ 	
\end{flushright}
\newpage

The non-perturbative calculation of the running coupling constant of 
QCD is certainly one very important problem. This program has been performed 
using the Schr\"odinger functional \cite{luscher}, the heavy quark potential
\cite{bali,bali2}, the Wilson loop \cite{lepage}, the Polyakov loop
\cite{divitis} and the three gluon coupling \cite{alles,alpha}. An usual approach to determine
the strong coupling constant (for instance in ref. \cite{lepage}) consists on the evaluation of 
an ultraviolet quantity and on the further comparison of the numerical data with perturbative
predictions. the gluon propagator at large momenta is a good candidate to be used.

Much work has been recently devoted to the study on the lattice of the gluon propagator 
\cite{bernard,marenzoni,leinweber,nakamura,nakajima,cucchieri,ma},
 the effort being mainly concentrated
on infrared behavior. The authors of \cite{leinweber} 
have started an ultraviolet study by comparing the asymptotic behavior
 to the one loop QCD prediction. In this
paper we would like to concentrate on the asymptotic  behavior of the lattice
propagator in the Landau gauge by a systematic use of the three-loops QCD
prediction, leaving a study of the infrared to a later publication.
 
The propagator is evaluated to a much better statistical accuracy than the
coupling constant computed in \cite{alles} and \cite{alpha}. The three-loops fit
should allow a determination of the strong coupling  constant and hence of 
$\Lambda_{\rm QCD}$
to a good accuracy in view of the very large statistics we have accumulated
(1000 configurations at $\beta=6.2$ on a $24^4$ lattice).
 This can be done in any renormalisation scheme as long as
 the anomalous dimension of the gluon field and the beta function are
known to three loop. 
A test of scheme dependence can thus be performed
rather extensively. It will turn out that although the statistical accuracy is
very good as expected, there remains a large systematic  error due to the non
asymptoticity of the gluon propagator. 

 In section 1, the general principle of the method is explained. In section 2 
  we perform a discussion of the lattice artifacts related to the hypercubic
  geometry, and we propose a solution to eliminate them. 
  In section 3 the fit is performed in the $\widetilde{\hbox{MOM}}$ scheme. 
  In section 4 we discuss in general the scheme dependence.
  We conclude in section 5.

\section{General description of the method}
\label{method}

The Euclidean two point Green function in momentum space writes in the
 Landau gauge: 
\beq
	G_{\mu_1\mu_2}^{(2)\,a_1 a_2}(p,-p)=G^{(2)}(p^2) 
	\delta_{a_1 a_2} \left(\delta_{\mu_1\mu_2}-
	\frac{p_{\mu_1}p_{\mu_2}}{p^2}\right)\label{G2}
\eeq
where $a_1, a_2$ are the color indices ranging from 1 to 8. 

In any regularization scheme (lattice, dimensional regularization, etc.)
with a cut-off $\Lambda$ ($a^{-1}$, $(d-4)^{-1}$)
the bare gluon propagator in the Landau gauge is such that
\beq
\lim_{\Lambda\to \infty}
\frac{d \ln [p^2 G^{(2)}_{\rm bare}(p^2,\Lambda)]}{d \ln p^2},
\label{obs}\eeq
  is independent of the
regularization scheme\footnote{We will return to 
this statement in the next subsection.}. Lattice calculations provide us with an 
evaluation of the bare propagator, and hence of the logarithmic derivative
in eq. (\ref{obs}). The latter is an observable
on which we will concentrate in this paper and
from which we will compute the strong coupling constant. 
 
In the MOM (or $\widetilde{\rm MOM}$) scheme,
calling $Z_3(q)$
the gluon renormalisation constant:
\beq
	Z_3(q,\Lambda)=  q^2 G^{(2)}_{\rm bare}(q^2,\Lambda)\label{Z3}
\eeq
where $q$ is a positive energy scale defined by  $q^2\equiv p^2$,
it happens that the expression in eq. (\ref{obs})
 is simply equal to 
$\lim_{\Lambda\to\infty}{d \ln Z_3(q,\Lambda)}/{d \ln q^2}$ .

An important point has to be stated here. The fact that $Z_3$ is the
renormalisation constant in the MOM scheme  does not
constrain us to stick to the MOM scheme all through. Equation 
(\ref{obs}) defines unambiguously a quantity, the evolution of which we
can study in {\sl any scheme}. This evolution is given by the anomalous
dimension of $Z_3$:

\beq
\frac{d \ln Z_3(q,\Lambda)}{d \ln q^2}=\Gamma(\alpha)=-\left( \frac{\gamma_0}
{4\pi} \alpha +\frac{\gamma_1} {(4\pi)^2} \alpha^2
+\frac{\gamma_2} {(4\pi)^3} \alpha^3 + O(\alpha^4)\right)\label{gamma}
\eeq
where it is understood that the coupling constant in a given scheme is a
function of $q$ such that
\beq
\frac{\partial \alpha}{\partial \ln q}=\beta(\alpha)= -\frac{\beta_0}
{2\pi} \alpha^2 -\frac{\beta_1} {(2\pi)^2} \alpha^3
-\frac{\beta_2} {(4\pi)^3} \alpha^4 + O(\alpha^5)\label{beta}
\eeq
with
\beq
\beta_0=11,\quad \beta_1=51,\quad \gamma_0=\frac {13}{2}\label{gamma0}
\eeq
in the flavorless case, $\gamma_1,\gamma_2$ and $\beta_2$ being scheme dependent. As we shall see
later, there is one scheme-independent relation between $\gamma_1,\gamma_2$ 
and $\beta_2$. To be specific, in the flavorless $\widetilde{\hbox{MOM}}$ 
scheme, for which $\beta_2$ is known
\cite{beta2}:
\beq
\beta_2\simeq 4824.,\quad\gamma_1= \frac {29}8,\quad \gamma_2 \simeq 960.
\label{beta2}
\eeq
The calculation of $\gamma_1$ and $\gamma_2$ will be explained later.  

From eqs.~(\ref{gamma}) and (\ref{beta}) it is easy to integrate
simultaneously, up to three loops, $\ln Z_3(q)$ and $\alpha(q)$ provided
one is given 
the values $Z_3(\mu)$ and $\alpha(\mu)$ at some initial value $q=\mu$,

\bea
q(\alpha) \ = \mu \left({\alpha\over \alpha(\mu)}\sqrt{{32 \pi^2 \beta_0 +16 \pi \beta_1
\alpha(\mu)+ \beta_2 \alpha^2(\mu)\over 32 \pi^2 \beta_0 +16 \pi \beta_1
\alpha+ \beta_2 \alpha^2}} \; \right)^{{\beta_1\over \beta_0^2}} \nonumber \\
\exp\left\{{2 \pi\over \beta_0}
\left({1\over \alpha}-{1\over \alpha(\mu)} \right) + {1 \over 2 \beta_0^2}\left(\beta_0
\beta_2-4 \beta_1^2\right) \left[H(\alpha)-H\left(\alpha(\mu)\right) \right] \right\} \ ,
\label{almu}
\eea

\noindent and

\bea
Z_3(\alpha)=Z_3(\mu) \left({\alpha\over \alpha(\mu)}\right)^{{\gamma_0\over 
\beta_0}} \left({32 \pi^2 \beta_0 +16 \pi \beta_1
\alpha+ \beta_2 \alpha^2 \over 32 \pi^2 \beta_0 +16 \pi \beta_1
\alpha(\mu)+ \beta_2 \alpha^2(\mu)} \right)^{{\gamma_2\over \beta_2}-{\gamma_0
\over 2 \beta_0}} \nonumber \\
\exp\left\{ \left(2\gamma_1-{2\beta_1\gamma_0 \over \beta_0}-4{\beta_1\gamma_2\over
\beta_2} \right) \left[H(\alpha)-H\left(\alpha(\mu)\right) \right] \right\}
\label{zmu} ;
\eea

\noindent where, for $2 \beta_0 \beta_2 - 4 \beta_1^2 > 0$~\footnote{We write the formal solution
of Eqs. (\ref{gamma},\ref{beta}) for the case of positive discriminant because this is the case,
for instance, of MOM and $\overline{{\rm MS}}$ schemes.},

\beq
H(\alpha)\ = \ {1 \over \sqrt{2 \beta_0 \beta_2 - 4
\beta_1^2}}\arctan \left( {1\over 4 \pi} {8 \pi \beta_1 + \beta_2 \alpha
\over \sqrt{2 \beta_0 \beta_2 - 4 \beta_1^2}}\right) \ .
\eeq

\noindent Eqs. (\ref{almu},\ref{zmu}) give a parametric representation of
the exact solution of the coupled differential equations (\ref{gamma}) and (\ref{beta}), 
where $\alpha$ can be considered just as a parameter connecting $Z_3$ and $q$.
The use of this parametric representation allows to work with exact solutions of Eqs.
(\ref{gamma}) and (\ref{beta}), the computation of $\alpha(q)$ from Eq. (\ref{almu}) being
only approximatively possible. However, as explained in section 4, no important discrepancy
comes from the consistent perturbative expansion on $\alpha$ of the exact solutions 
(\ref{almu},\ref{zmu}).

{\sl Our general ``strategy'' will be to  fit the lattice results with one 
solution (\ref{almu},\ref{zmu})}.
If an acceptable best fit exists, it will prove that the lattice results do
agree in the considered energy range with three-loops perturbative QCD. 
Moreover, it will provide us with the best initial values $Z(\mu)$ and
$\alpha(\mu)$. $Z(\mu)$ is just an overall multiplicative constant, but
the knowledge of $\alpha(\mu)$ in a given scheme allows to compute an
 estimate of
$\Lambda_{\rm QCD}$ in this scheme, and hence $\Lambda_{\overline {\rm MS}}$.

To our knowledge, this method of determining the strong coupling constant on the
lattice is new. 
In \cite{leinweber}, the asymptotic behaviour of the gluon propagator has been
compared to the one loop perturbative QCD prediction. The authors indeed find
rough agreement. However, the value of $\Lambda$ that one can extract from the
one loop fit is not meaningful since, $\gamma_0$ being scheme independent,
 one does not know in which scheme it is computed. In the following we will
 systematically compare the two loops to the three loops result in a large set
 of schemes, and we shall
 conclude that at the accessible scales of 2.5 - 5. GeV, the gluon propagator 
 does not scale to two loops and hardly does to three loops in the most
 favorable schemes. In the $\overline{\rm MS}$ scheme, no sign of scaling is found. 
 
\subsection{Three loops expansion}

In a general renormalisation scheme, which we call SC, and in a fixed gauge
\beq
 G^{(2)}_{\rm bare}(p^2,\Lambda) = Z_{3\, {\rm SC}}(q, \Lambda)
[G^{(2)}_{\rm SC}(p^2,q)+O(1/\Lambda)]\label{general}
\eeq
where $Z_{3\, {\rm SC}}$ is the renormalisation constant in the scheme
 and  $G^{(2)}_{\rm SC}$ the renormalized gluon propagator.
Notice by the way, that from eq. (\ref{general}) one immediately 
obtains the well
known above-mentioned result that 
\beq
\lim_{\Lambda\to\infty}\frac{d \ln Z_{3\, {\rm SC}}(q,\Lambda)}
{d \ln q^2}=-
\frac{d \ln G^{(2)}_{\rm SC}(p^2,q)}{d \ln q^2}
\eeq
 is finite and independent of the regularization scheme, while the r.h.s is
 independent of $p^2$. From (\ref{Z3}) and
 (\ref{general})
 \beq
 Z_3(q,\Lambda)=  q^2
 G^{(2)}_{\rm bare}(q^2,\Lambda) = Z_{3\, {\rm SC}}(q, \Lambda)
 [q^2 G^{(2)}_{\rm SC}(q ^2,q)+O(1/\Lambda)] \label{curly}
 \eeq
whence
\beq
\lim_{\Lambda\to\infty}\frac{d \ln Z_3(q,\Lambda)}{d \ln q^2}= 
\lim_{\Lambda\to\infty}\frac{d \ln Z_{3\, {\rm SC}}(q,\Lambda)}
{d \ln q^2}
+\frac{d \ln [q^2\, G^{(2)}_{\rm SC}(q^2,q)] }{d \ln q^2}
\label{logZ3}\eeq

 Expanding  eq. (\ref{logZ3})  in $\alpha$
introduces a dependence on the scheme in which
$\alpha$ is expressed.
We will now specify the scheme SC to be the $\overline {\rm MS}$ scheme since
the gluon propagator anomalous dimension has been computed to three loops
(see eq. (8) in \cite{larin}):
\beq
\lim_{\Lambda\to\infty}\frac{d \ln Z_{3\,\overline {\rm MS}}(q,\Lambda)}
{d \ln q^2}\simeq -\frac{13}
{2(4\pi)} \bar\alpha -\frac{531} {8(4\pi)^2} \bar\alpha^2
-\frac{29311.} {32(4\pi)^3} \bar\alpha^3 + O(\bar\alpha^4)
\label{gammaMS}
\eeq
where $\bar \alpha$ is the coupling constant in the $\overline {\rm MS}$
scheme and where the $\epsilon\equiv d-4\to 0$ limit has been taken. 
Using now\footnote{$[q^2\, G^{(2)}_{\overline{\rm MS}}(q^2,q)]^{-1}$
is equal to $J^{\rm ren}$ as given in
 eq. (8.13) in \cite{davyd}.}
the expression for  $q^2\, G^{(2)}_{\overline{\rm MS}}(q^2,q)$ in \cite{davyd}
   we can rewrite eq. (\ref{logZ3}) as

\[
\lim_{\Lambda\to\infty}\frac{d \ln Z_3(q,\Lambda)}{d \ln q^2}\equiv
\Gamma(\bar\alpha)=
 -\frac{\gamma_0}
{(4\pi)} \bar\alpha -\frac{\overline\gamma_1} {(4\pi)^2} \bar\alpha^2
-\frac{\overline\gamma_2} {(4\pi)^3} \bar\alpha^3 + O(\bar\alpha^4)
\]
\beq
\simeq -\frac{13}
{2(4\pi)} \bar\alpha -\frac{155.3} {(4\pi)^2} \bar\alpha^2
-\frac{6656.} {(4\pi)^3} \bar\alpha^3 + O(\bar\alpha^4)
\label{gMS}
\eeq

\subsection{General three-loops schemes}

If the strong coupling constant is expressed in some other scheme, 
which we shall call generically the ``tilde'' scheme,
 once known $\widetilde \alpha$
as a function of $\bar \alpha$,  changing scheme simply amounts to a change of
variables. If
\beq
\widetilde \alpha = \bar \alpha + \frac a {4\pi} \bar \alpha^2
+ \frac b {(4\pi)^2} \bar \alpha^3 + \dots \label{cdv}
\eeq
then as shown for example in \cite{beta2},

\beq
 \Lambda_{\overline{\rm MS}} = \widetilde \Lambda e^{-\frac a {2 \beta_0}},\quad
 \widetilde \beta_2=\beta_{2\,\overline{\rm MS}} + 2 \beta_0 (b-a^2)-4a\beta_1
 \; .
 \label{deuxlambdas}
\eeq
The $\widetilde\gamma_i$'s being defined from
\beq
 \lim_{\Lambda\to\infty}\frac{d \ln Z_3(q,\Lambda)}{d \ln q^2}=
\Gamma(\widetilde\alpha)\simeq -\frac{\gamma_0}
{(4\pi)} \widetilde\alpha -\frac{\widetilde\gamma_1} {(4\pi)^2} \widetilde\alpha^2
-\frac{\widetilde\gamma_2} {(4\pi)^3} \widetilde\alpha^3 + O(\widetilde\alpha^4),
\label{gtilde}
\eeq
 are then computed from the change of variables (\ref{cdv}) inserted
into (\ref{gMS}), 
finally leading to the relations

\beq
 \Lambda_{\overline{\rm MS}} = \widetilde \Lambda \,\exp\left[{\frac
 {\widetilde\gamma_1-\overline\gamma_1} 
 {2 \gamma_0\beta_0}}\right],\label{2lambdas}
 \eeq
 \beq
 \widetilde \beta_2=\beta_{2\,\overline{\rm MS}} + 2 \beta_0 \left[\frac
 {\overline\gamma_2-\widetilde\gamma_2}{\gamma_0}+
 \left(\frac{\widetilde\gamma_1}{\gamma_0}\right)^2-
 \left(\frac{\overline\gamma_1}{\gamma_0}\right)^2\right]
 -4\beta_1\left[\frac
 {\overline\gamma_1-\widetilde\gamma_1}{\gamma_0}\right]
 \label{beta2tilde}
\eeq

It results that $\widetilde \beta_2$, $\widetilde\gamma_1$, and $\widetilde\gamma_2$
are not independent parameters. Knowing two among these three parameters allows
the computation of the third one. We stress that in any ``tilde'' scheme, only 
two parameters 
are required for any purpose we need in the study that we present in this paper.
In other words, the set of renormalisation schemes 
is a two parameter space in which the $\overline {\rm MS}$ is 
one point, and $\widetilde {\rm MOM}$  another.
On the following, we will use 
the coordinates ($\widetilde\gamma_1,\widetilde\gamma_2$) to characterize any renormalisation
scheme of the two parameter space.
The possible use of this large parameter space of schemes will provide us 
with a tool which we will exploit extensively in the following. Note that
when $\tilde\gamma_1=\tilde\gamma_2=0$, the coupling constant
$\widetilde\alpha$ is nothing but the ``effective charge'', \cite{grunberg},
 associated to the observable defined in eq. (\ref{obs}). In reference
 \cite{grunberg} this charge was argued to be the proper object on which
  perturbation theory
 applies. We will generalise the latter philosophy by considering also
 schemes with a non-vanishing  $\widetilde\gamma_1,\widetilde\gamma_2$.

In any ``tilde'' scheme, once evaluated $\widetilde \alpha(\mu)$, we can compute
 $\widetilde \Lambda$ up to three loops. For a discussion of the different formulae,
 see \cite{beta2}. At two loops we will use the conventional formula
 \beq       
             \widetilde \Lambda^{(c)} \equiv \mu \exp\left (\frac{-2 \pi}{\beta_0
	      \widetilde\alpha(\mu^2)}\right)\times
	      \left(\frac{\beta_0  \widetilde\alpha(\mu^2)}{4 \pi}\right)^{-\frac {\beta_1}
	      { \beta_0^2}}\label{lambda}
\eeq

At three loops,   
 we will use, when $\Delta\equiv 2\beta_0\widetilde\beta_2-4\beta_1^2>0$,
either the unexpanded formula
\[
 \widetilde\Lambda^{(3)}\equiv\widetilde\Lambda^{(c)}(\widetilde\alpha)\left(1+\frac 
 {\beta_1\widetilde\alpha}{2\pi\beta_0}+
 \frac
 {\widetilde\beta_2\widetilde\alpha^2}{32\pi^2\beta_0}\right)^{\frac{\beta_1}{2\beta_0^2}}
  \]
  \beq \exp
\left\{\frac{\beta_0\widetilde\beta_2-4\beta_1^2}{2\beta_0^2\sqrt{\Delta}}\left[
\arctan\left(\frac{\sqrt{\Delta}}{2\beta_1+\widetilde\beta_2\widetilde\alpha/4\pi}\right)
-\arctan\left(\frac{\sqrt{\Delta}}{2\beta_1}\right)\right]\right\}
 \label{lambda3}\eeq
  or the expanded one:
 \beq
 \widetilde\Lambda^{(3)}_{\rm exp}\equiv\widetilde\Lambda^{(c)}(\widetilde\alpha)\left(1+\frac 
 {8\beta_1^2-\beta_0\widetilde\beta_2}{16\pi^2\beta_0^3}\widetilde\alpha
 \right)
  \label{lambdaexp}\eeq
In eqs (\ref{lambda3}-\ref{lambdaexp}) we have omitted for simplicity to write 
the $\mu^2$ dependence of $\widetilde\alpha$.
 
\section{Hypercubic artifacts and other $O( \lowercase{a}^2 \lowercase{p}^2)$ effects}

We refer to \cite{alpha} for a description of the lattice simulations which have
been performed, the calculation of the Green functions, their Fourier transform, the
checks of the $\delta_{a_1,a_2}$ color dependence of the propagators,
and the set of momenta considered for the different lattices studied. We will
use in this study 1000 configurations at $\beta=6.2$ on a $24^4$ lattice and 
100 configurations at $\beta=6.4$ on a $32^4$ lattice. The large statistics
involved will reduce the statistical error to a negligible value.

\begin{figure}[hbt]
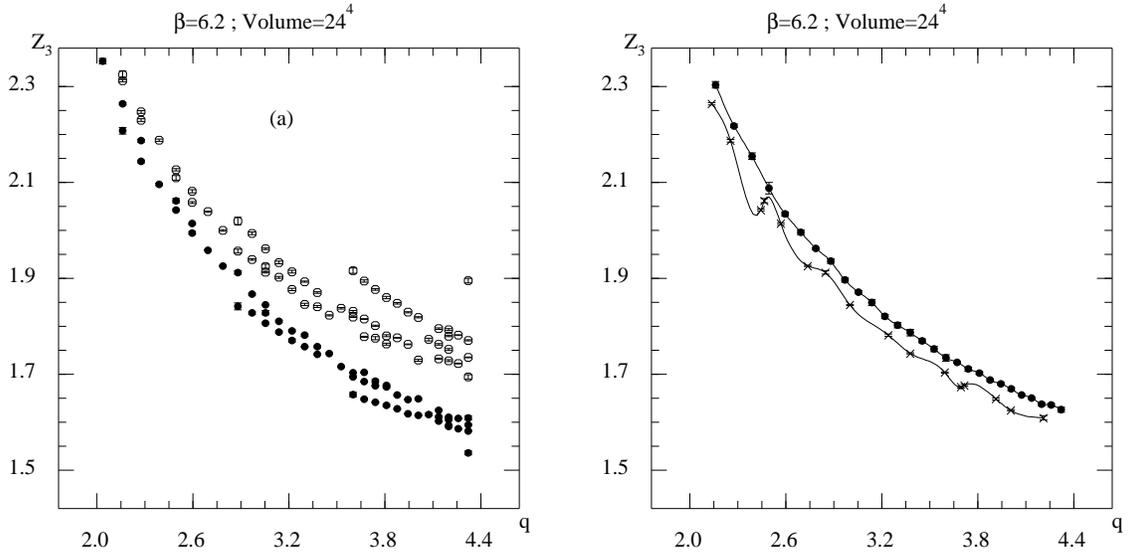

\begin{center}
\leavevmode
\mbox{\epsfig{file=ZAtot_orbites.eps,height=8cm}}
\mbox{\epsfig{file=ZAcompare.eps,height=8cm}}
\end{center}
\caption{\it Plot (a) shows  (open circles) $Z_3(q)$, eq. (\ref{Z3}), as a function of the scale
 $q$ with
 a distinct point for each orbit. It also shows (black circles) $Z_3(q)$ with
 the factor $q^2$ replaced by $\tilde q^2$ in eq. (\ref{Z3}). Even with the latter
 which reduces somehow the dispersion, 
  the difference between the individual orbits
at a same $q$ exceeds by far the statistical errors. Plot (b)
compares the ``democratic'' selection among orbits (lower curve), as a
function of $\widetilde q$ and the curve extrapolated to $p^{[4]}=0$ as 
 in eq. (\ref{roiesnel}). The latter procedure gives obviously a much 
 smoother result.}
\label{a2}
\end{figure}

In a finite hypercubic volume the momenta are the discrete sets of vectors 
\beq
p_\mu=\frac{2\pi}L n_\mu \label{momentum}
\eeq

\noindent where $n_\mu$ are integer  and $L$ is the lattice size. As explained in 
\cite{alpha}, we have averaged the propagators on the hypercubic isometry group
$H_4$. The momenta corresponding to {\it e.g} $n_\mu=(2,0,0,0)$
and $n_\mu=(1,1,1,1)$, belong to different orbits although they both have the same
$p^2=4 (2\pi)^2/L^2$, {\it i.e.} they belong to the same orbit of the {\sl continuum}
isometry group $SO(4)$. Our statistical errors are so tiny that the difference 
between the evaluated propagators for two such orbits of same $p^2$ are quite
visible, as shown in fig. \ref{a2}(a). Such differences are understood 
 as an $O(a^2p^2)$ artifact of the
 hypercubic geometry of the lattice. For example, if one  uses,
  rather than (\ref{momentum}), the momentum 
  
 \beq 
 \widetilde p_\mu = \frac 2 a \sin\left(\frac{a p_\mu}{2}\right).\label{ptilde}
\eeq 
the resulting momentum squared differs by a relative $O(a^2p^2)$:
\beq
\widetilde p^2=p^2 - \frac 1 {12} a^2 p^{[4]} + \dots,\quad {\rm where}\quad
p^{[4]}\equiv \sum_\mu p_\mu^4
\eeq

\noindent To reduce the hypercubic artifact, the authors of \cite{leinweber} advocate both
the use of (\ref{ptilde}) rather than (\ref{momentum}), and the restriction to a
``democratic'' set of momenta, i.e. those for which $p^2$ 
is rather equally distributed on the different directions ($p_\mu^2$).

\begin{figure}[hbt]
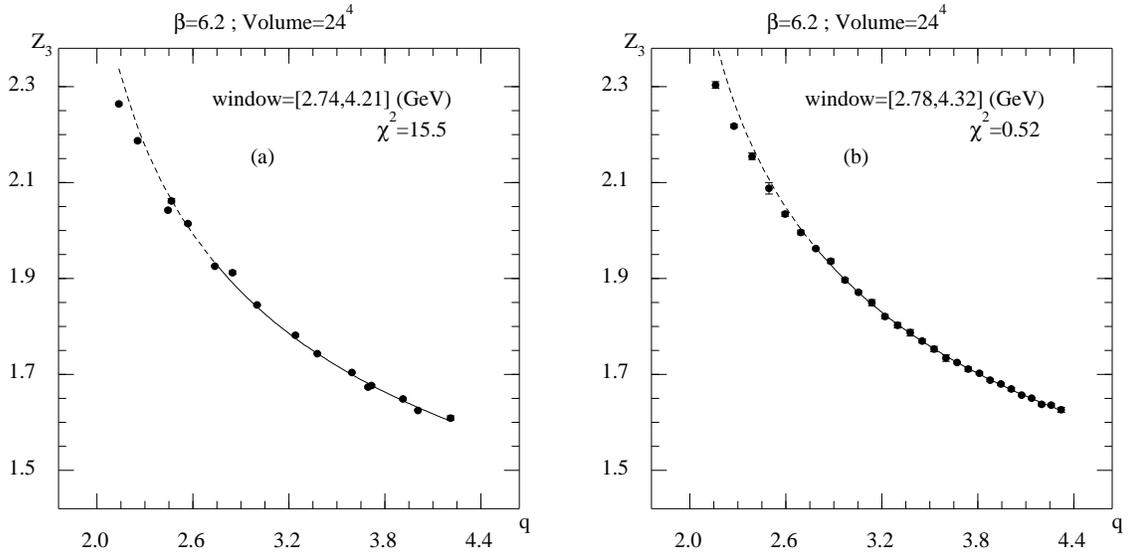

\begin{center}
\leavevmode
\mbox{\epsfig{file=ZAfit_dem.eps,height=8cm}}
\mbox{\epsfig{file=ZAfit_med.eps,height=8cm}}
\end{center}
\caption{\it Plot (a) shows (full line) the best fit of the three loops 
formula to $G_2(\widetilde q^2)\widetilde q^2$ with the ``democratic'' selection of orbits 
for 2.74 GeV $\le \widetilde q \le $ 4.21 GeV. The fit is prolongated 
outside the energy window in dashed line.
The $\chi^2$ per d.o.f is 15.5.
Plot (b) shows the best fit of $G_2(q^2) q^2$ extrapolated to $p^{[4]}=0$
according to eq. (\ref{roiesnel}) for 
2.78 GeV $\le \widetilde q \le $ 4.32 GeV. The $\chi^2$ per d.o.f is 0.52.
The improvement of the $\chi^2$ per d.o.f by a factor $\sim$ 30
is dramatic. Notice furthermore that the number of points is larger in plot (b)
because the ``democratic'' selection eliminates all the orbits for some values
of $q$.}
\label{a2fit}
\end{figure}

We will now propose another approach which we will eventually compare to the
``democratic'' one. In general, different orbits with the same $p^2$ have
different $p^{[4]}$. For example (2,0,0,0) has $n^{[4]}\equiv \sum n_\mu^4 = 16$ while 
(1,1,1,1) has $n_4=4$. On the lattice, the function $G^{(2)}(p^2)$ defined in
(\ref{G2}) is indeed a scalar form factor invariant under $H_4$ that we shall
assume to be a smooth function. The general structure of polynomials invariant 
under a finite group is known from invariant theory \cite{inv}. For our purpose
it is sufficient to know that any smooth $H_4$-invariant function is 
indeed a function of the 4 invariants $p^{[n]}=\sum_{\mu}p_{\mu}^n,\, n=2,4,6,8$.
 We will
neglect the invariants with degree higher than 4 since they vanish at least as
$a^4$ and parametrize the lattice two-point scalar form factor as a function
 $G_{\rm lat}^{(2)}(p^2, p^{[4]})$. When several orbits exist for 
 one $p^2$, it is 
possible to extrapolate to $p^{[4]}=0$ and we define

\beq
G^{(2)}_{\rm bare}(p^2) \equiv  \lim_{p^{[4]}\to 0}
G_{\rm lat}^{(2)}(p^2,p^{[4]})\label{roiesnel}
\eeq
It is easy to see that, neglecting $O(a^4)$, if one uses (\ref{ptilde}) instead of 
(\ref{momentum}), defining 
accordingly $\widetilde p^{[4]}$   and
$\widetilde G_{\rm lat}^{(2)}(\widetilde p^2,\widetilde p^{[4]})$, one has
\beq
\lim_{p^{[4]}\to 0} G_{\rm lat}^{(2)}(p^2,p^{[4]}) = 
\lim_{\widetilde p^{[4]}\to 0}\widetilde  G_{\rm lat}^{(2)}(\widetilde p^2,\widetilde p^{[4]})
\label{egalite}\eeq
where the limit in the r.h.s. is taken at constant $p^2$.
We have indeed numerically checked form eq. (\ref{egalite})
the absence of sizable $O(a^4)$ effects.

In fig. \ref{a2}(b), the ``democratic'' $Z_3(\widetilde q)$ and the one computed from
eq. (\ref{roiesnel}) are compared. The latter provides {\sl a much smoother
$Z_3(q)$}. This is confirmed by the $\chi^2$ of the fit in the next
section. The best fit of  the ``democratic'' $Z_3$ using the eqs. (\ref{gamma}) 
and (\ref{beta}) gives a $\chi^2$ per d.o.f of 15.5, see fig \ref{a2fit}(a),
while the best fit to
$Z_3$ computed from eq. (\ref{roiesnel}) gives a $\chi^2$ per d.o.f of
$\simeq$ 0.52, fig. \ref{a2fit}(b).
For sure, the $p^{[4]}\to 0$ extrapolation increases the errors on $Z_3$ as
compared to the errors in the individual orbits, but not too much. We have
computed the errors on the extrapolated points using the jackknife method.
 Typically
the extrapolation increases the errors by a factor $\sim 2$ which 
cannot account for an improvement by a factor $\sim 30$
on the $\chi^2$ per d.o.f. While this paper was in writing appeared 
a study by J.P. Ma, \cite{ma}, who suggests the use
of the variable $\widetilde p^2 + a^2 \widetilde p^{[4]}/12$
to cure hypercubic artifacts.
 He shows a significant smoothing (fig. 2 in ref. \cite{ma}).

We therefore conclude that eq. (\ref{roiesnel}) allows to eliminate in a
consistent way the hypercubic artifacts. This does not mean that 
all $O(a^2)$ artifacts are thus eliminated: for instance, the 
lattice artifacts $\propto a^2(p^2)^2$, which do not break $SO(4)$ invariance,
 are obviously not.
Only a comparison of our results for different lattice spacings 
 allows to estimate the latter effects. 

\section{General fit in the  $\widetilde {\rm MOM}$ scheme}

We first show the fit of our data at $\beta=6.2$ (1000 configurations)
with solutions of the coupled 
differential equations  (\ref{gamma}) and (\ref{beta}) in the 
$\widetilde {\rm MOM}$ scheme.
 It has been performed on the lattice 
propagator in the energy window\footnote{This best fit has been
 performed on a slightly different energy window than
 in fig. \ref{a2fit}(b).} 2.97 - 4.32 GeV. The result of the fit is
\beq
Z_3(4.32\ {\rm GeV})=1.625(5),\quad \alpha_{\widetilde{\rm MOM}}(4.32\ {\rm GeV})=
 0.3005(15),\quad \chi^2/{\rm d.o.f.} = 0.44 \label{resmom}
\eeq

The $\chi^2$ per d.o.f. is significantly smaller than 1. This
feature may be a sign of some correlation between the points at different values of
the energy $q$. Using (\ref{lambda3}-\ref{lambda}) and (\ref{2lambdas}) we obtain
\beq
\Lambda^{(3)}_{\overline {\rm MS}}\simeq 0.346\; 
\Lambda^{(3)}_{\widetilde {\rm MOM}} = 354.2 \pm 2.5 \ {\rm MeV},\quad
\Lambda^{(c)}_{\overline {\rm MS}}\simeq 0.346\; 
\Lambda^{(c)}_{\widetilde {\rm MOM}} \simeq 454  \ {\rm MeV},
\label{lambdamom}
\eeq
where the error is only statistical. The difference between the
two-loops and the three-loops result is large. We will discuss this
 feature in the next section.
 We have checked that
the result is not significantly changed when other energy windows 
 are  used.

The study of our data at $\beta=6.4$ (100 configurations) will be useful to estimate 
lattice artifacts $\propto a^2(p^2)^2$. 
The best fit is now obtained for the window 2.97 - 5.48 GeV . The results 
for such a fit are the following

\beq
Z_3(5.48\ {\rm GeV})=1.452(18),\quad \alpha_{\widetilde{\rm MOM}}(5.48\ {\rm GeV})=
 0.255(4),\quad \chi^2/{\rm d.o.f.} = 1.1 \label{resmom6.4}
\eeq
For comparison with (\ref{resmom}), performing the fit in the energy window
2.97 - 4.32 GeV, we obtain $\alpha_{\widetilde{\rm MOM}}
(4.32\ {\rm GeV})= 0.287(10)$. 
\noindent Notice that the $p^{[4]}\to0$ extrapolation 
meant to reduce hypercubic artifacts,
 eq. (\ref{roiesnel}), 
is slightly less efficient than at $\beta = 6.2$, maybe due to the smaller 
statistics.  Still it allows a gain of a factor
 15 for the $\chi^2$ with regard to  the ``democratic" analysis. It results

\beq
\Lambda^{(3)}_{\overline {\rm MS}}\simeq 0.346\, 
\Lambda^{(3)}_{\widetilde {\rm MOM}} = 346 \pm 8 \ {\rm MeV},\quad
\Lambda^{(c)}_{\overline {\rm MS}}\simeq 0.346\, 
\Lambda^{(c)}_{\widetilde {\rm MOM}} \simeq 441 \ {\rm MeV}. \label{lambdamom6.4}
\eeq

\noindent The comparison between the results at $\beta = 6.4$ and at 
$\beta = 6.2$ seems to indicate that the 
lattice artifacts other than those $\propto a^2 p^{[4]}$ are not too important.
We did not use our data at $\beta=6.0$ since they do not reach a large enough
energy scale.

\section{Scheme dependence}

\begin{figure}[hbt]
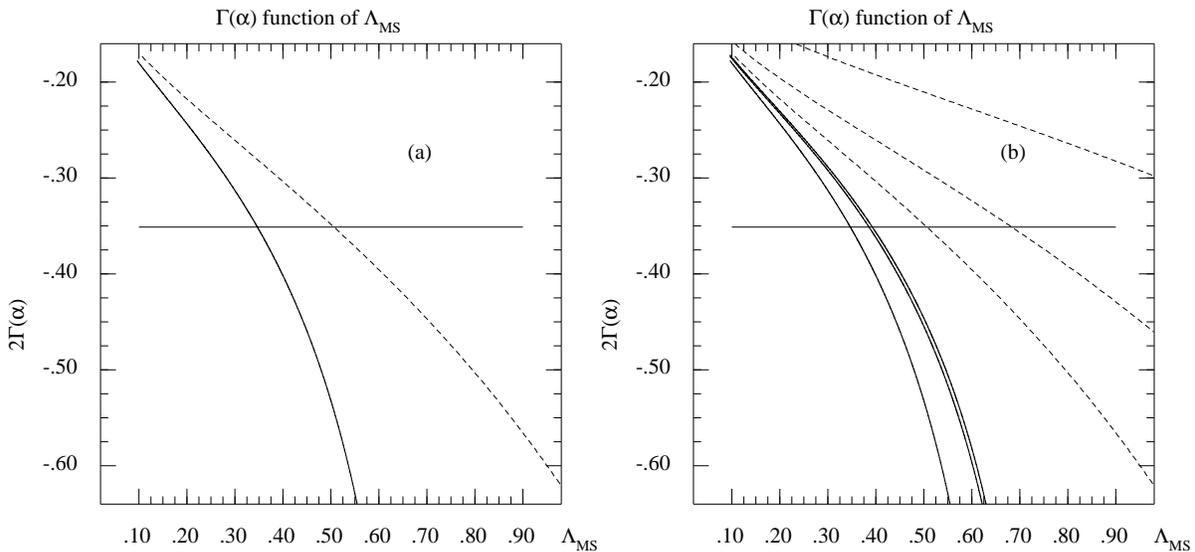

\begin{center}
\leavevmode
\mbox{\epsfig{file=pente_lambda.eps,height=8cm}}
\mbox{\epsfig{file=p_de_l_all.eps,height=8cm}}
\end{center}
\caption{\it Plot (a) shows in full line $2 \Gamma(\alpha)$ in the $\widetilde {\rm
MOM}$
scheme as a function of  $\Lambda^{(3)}_{\overline {\rm MS}}$, using eqs (\ref{gamma}) and
  (\ref{lambda3}). It shows in dashed line $2 \Gamma(\bar\alpha)$ 
  in the $\overline {\rm MS}$
scheme as a function of  $\Lambda^{(3)}_{\overline {\rm MS}}$, using eqs 
(\ref{gMS}) and (\ref{lambda3}). The horizontal line corresponds to a typical
value of the logarithmic derivative ${\partial \ln Z_3(q)}/{\partial \ln q}$ 
  evaluated from our lattice data around 4 GeV. 
Plot (b) adds to the preceding plot the curves  $2 \Gamma(\alpha)$
(resp. $2 \Gamma(\bar\alpha)$) truncated
at first order i.e. $\gamma_1=\gamma_2=0$ (resp. $\bar\gamma_1=\bar\gamma_2=0$)
and at second order i.e. $\gamma_2=0$ (resp. $\bar\gamma_2=0$) in full lines (resp.
dashed lines). In both cases the curves range  from right to left
with increasing number of terms. In all cases an energy scale of 4.1 GeV 
has been used to compute $\Lambda^{(3)}_{\overline {\rm MS}}$.   
}
\label{plot-gamma}
\end{figure}

\subsection{$\overline {\rm MS}$ scheme}

It might look a little too involved to fit the data as a function of
$\alpha_{\widetilde{\rm MOM}}(\mu)$ and then to convert 
$\Lambda^{(3)}_{\widetilde {\rm MOM}}$ into
$\Lambda^{(3)}_{\overline {\rm MS}}$. Would it not be simpler to work all the
way in $\overline {\rm MS}$, i.e. to perform the fit as a function of $\bar
\alpha(\mu)$ and to convert directly the result into  
$\Lambda^{(3)}_{\overline {\rm MS}}$ ?
We have done this exercise and found  

\beq
Z_3(4.3 {\rm GeV})=1.623(5),\quad \bar\alpha(4.3 {\rm GeV})=0.207(2)
,\quad \chi^2/{\rm d.o.f.} = 0.64 \label{resms}
\eeq
leading via eq.  (\ref{lambda3}) to
\beq
\Lambda^{(3)}_{\overline {\rm MS}} = 549 \pm 9 \ {\rm MeV},\quad
\Lambda^{(c)}_{\overline {\rm MS}} \simeq  765\ {\rm MeV}\label{lambdams}
\eeq

The discrepancy between (\ref{lambdams}) and (\ref{lambdamom}) comes as a big
 surprise, since we have only performed a change of variables. The only
difference between the two methods comes from the truncation of the perturbative
series. To be more precise, when we work with $\alpha_{\widetilde{\rm MOM}}$
to three loops, we truncate the series (\ref{gamma}) and (\ref{beta}) beyond the
third term. Changing variables to $\bar \alpha$ means expanding $\bar\alpha$
in terms of $\alpha_{\widetilde{\rm MOM}}$ up to the third term, (\ref{cdv}),
 implementing this change into the functions $\Gamma(\alpha)$ and
 $\beta(\alpha)$ and expanding the result up to the third term. This final
 expansion rejects in a consistent way some parts which are formally 
 equivalent to four and higher 
 loops. Could it be that four-loops effects 
 are so large as to induce the discrepancy between (\ref{lambdams}) and  
 (\ref{lambdamom}), although we are working at an energy scale larger than
  4~GeV~?
 
  To dig into this question, we have performed the following exercise. In
  fig. \ref{plot-gamma}(a) we plot $2\Gamma(\alpha)$ as a function of 
  $\Lambda^{(3)}_{\overline {\rm MS}}$, using eqs (\ref{gamma}) and
  (\ref{lambda3}). We do this in both schemes. The horizontal line, at 
  -0.35 is about the slope ${\partial \ln Z_3(q)}/{\partial \ln q}$ 
  evaluated from our lattice data around 4. GeV. What happens is now obvious. 
  the curve corresponding to $\overline {\rm MS}$ at three loops 
  is far from the $\widetilde {\rm MOM}$ one at such a large absolute value of
  the logarithmic slope as 0.35. It is also obvious from the fig.
  \ref{plot-gamma}(a) that, 
  had the slope been smaller in absolute value, i.e. had the horizontal line
  been, say at -0.20, the two 
  schemes would have agreed much better. But the lattice results show that such
  a slope can only be reached at a much larger energy than 4. GeV 
  even though we are working at three loops !! In other
  words, the lesson is that the Landau gauge gluon propagator reaches asymptotic scaling
  only at very large energies. We are indeed
  starting a study at larger energies.

  The next question to address is whether it is possible to choose the best
  scheme, in the sense of the scheme which is closest to asymptotics ? Of course
  we have a prejudice, from other theoretical and experimental 
 sources   about $\Lambda_{\overline {\rm MS}}$, including our own analysis of
 $\alpha$, \cite{alpha,beta2}, 
 that the result in
 (\ref{lambdamom}) is better. However, it would be more consistent to conclude about this 
 question with the only evidences
 coming from the analysis of the gluon propagator.
 In fig. \ref{plot-gamma}(b) we plot the same curve as in 
 fig. \ref{plot-gamma}(a), together
 with the equivalent plots, but taking $\gamma_2=0$, and the ones with
 $\gamma_1=\gamma_2=0$\footnote{We have kept for these curves the 
 three-loops computation of $\Lambda$ according to (\ref{lambda3}) with 
 non-vanishing $\gamma_2$:
 these are not genuine two or one-loop calculations.
  Later we will rather
 compare the three-loop result with the consistent 
 two-loop calculation of the same scheme, using (\ref{lambda}). Our qualitative
 conclusion about the ``best'' schemes will not change.}. It is clear that 
 the three curves are much closer in the 
 $\widetilde{\rm MOM}$ scheme  than in the $\overline {\rm MS}$ one. This 
 is an argument in favor of $\widetilde{\rm MOM}$ while 
 $\overline {\rm MS}$ proves really bad for this problem: the two-loops is
 more than 200 MeV away from the three-loops, not to speak of the one-loop which 
 does not even cross the line in our plot ! 
  
 \subsection{Search in the scheme space} 
 
 As long as we are engaged in comparing the schemes, why not exploit 
 the richness of the large scheme space to which we can access up to
  three loops simply by defining arbitrary couple of parameters
  ($\widetilde \gamma_1,
  \widetilde\gamma_2$). From (\ref{beta2tilde}) one also knows $\widetilde\beta_2$
  and from (\ref{2lambdas}) the ratio $\Lambda_{\overline{\rm
  MS}}/\widetilde\Lambda$. Hoping to find schemes for which asymptotic scaling
   is reached, 
  our strategy will be to scan the latter parameter space,
  and to try a fit of the data at $\beta=6.2$ in order to find the true
  asymptotic $\Lambda_{\overline{\rm MS}}$. We select the 
  domain of the ``good schemes'' by
  imposing some constraints which will be discussed right now, and the variation
  of $\widetilde \Lambda^{(3)}$ in the latter domain will provide an estimate of one
  source of systematic uncertainty. 
  
  In order to determine the good schemes we will impose three criteria. 
  The first two are the following:
  \beq |\Lambda^{(2)}_{\overline{\rm MS}}-\Lambda^{(3)}_{\overline{\rm MS}}|
  < \Delta_{\rm max},\quad
\delta\Lambda<\delta_{\rm max}\label{crit}
\eeq 
 with $\Lambda^{(3)}_{\overline{\rm MS}}$ computed 
to three loops, using (\ref{lambda3}), and $\Lambda^{(2)}_{\overline{\rm MS}}$
 computed to two
loops, i.e. taking the same $\widetilde\gamma_1$, but with 
$\widetilde\gamma_2=0$, and
(\ref{lambda}); $\delta\Lambda$ is the difference
$|\Lambda^{(3)}_{\overline{\rm MS}\,{\rm exp}}-
\Lambda^{(3)}_{\overline{\rm MS}}|$, with 
$\Lambda^{(3)}_{\overline{\rm MS}\,{\rm exp}}$ computed 
to three loops, using 
(\ref{lambdaexp}). Typically $\Delta_{\rm max}$ will be taken around 100 MeV
and $\delta_{\rm max}$ around 10 to 25 MeV. The reason for these selection
criteria should be rather clear: we wish the series of the functions 
$\beta(\widetilde\alpha)$ and
$\Gamma(\widetilde\alpha)$ to look like being in their convergence domain;
the difference between the two-loops and the three-loops should thus not  be
exceedingly large as discussed in the preceding subsection. Finally we wish 
that expanding the formula for $\Lambda^{(3)}$ does not change drastically
the result in order to keep some faith in the perturbative character of the
latter formula.

The criteria in (\ref{crit}) are not restrictive enough to forbid
exceedingly large values of $\widetilde\gamma_1$ and $\widetilde\gamma_2$.
To cure this we try some reasonable guesses about higher order
terms. We extend to four loops the method which lead us to 
 (\ref{beta2}),
 
\[ 
\widetilde\beta_3= \beta_{3} 
+{2\beta_{2}(\widetilde\gamma_1-\gamma_{1})\over \gamma_0}+{4\beta_{1}
(\widetilde\gamma_1-\gamma_{1})^2\over \gamma_0^2} 
-{4\beta_{0}(\widetilde\gamma_1-\gamma_{1})^3\over \gamma_0^3}
\]
\[
+{8\beta_{0}(\widetilde\gamma_1-\gamma_{1})(2\gamma_{1}^2-2\gamma_{1}
\widetilde\gamma_1-\gamma_0 \gamma_{2}+\gamma_0\widetilde\gamma_2)\over \gamma_0^3} 
\]
\[
-{4\beta_{0}\over\gamma_0^3}( -5\gamma_{1}^3+6\gamma_{1}^2\widetilde\gamma_1-
\gamma_{1}\widetilde\gamma_1^2 
+5\gamma_0\gamma_{1}\gamma_{2}-3\gamma_0\widetilde\gamma_1\gamma_{2}
-2\gamma_0\gamma_{1}\widetilde\gamma_2)\]
\beq
-\frac{4\beta_0(\widetilde \gamma_3-\gamma_3)}{\gamma_0}
 \label{beta3} 
\eeq
\noindent 
where $\gamma_i$ and $\beta_i$ are in the $\widetilde {\rm MOM}$ scheme, eq. 
(\ref{beta2}),  and where
the beta function is defined up to four loops as
\beq
\beta(\alpha)=-\frac{\beta_0}
{2\pi} \alpha^2 -\frac{\beta_1} {(2\pi)^2} \alpha^3
-\frac{\beta_2} {(4\pi)^3} \alpha^4 -\frac{\beta_3} {(4\pi)^4} \alpha^5 \;\;\; .
\eeq
Since $\gamma_3$ is unknown, we restrict the ``tilde'' scheme by imposing
 $\widetilde\gamma_3\equiv\gamma_3$.
We do not know $\beta_3$ in the $\widetilde {\rm MOM}$ scheme, but we 
believe that it cannot be too large because the good fit of $\alpha$, extracted
from the three gluon vertex in \cite{beta2}, using the 
 three-loops expression (i.e. with $\beta_3=0$), persisted down to rather low 
 energies.
Therefore, by bounding the distance between $\beta_3$ and $\widetilde\beta_3$
we will keep $\widetilde\beta_3$ reasonably small.
Now we are ready to add to (\ref{crit}) a third criterion
to bound a reasonable domain of the ``good schemes'': 

\beq
| \widetilde\beta_3 -\beta_3 | < 4 \pi \beta_{2}\label{crit2}
\eeq
It is worth noticing that this last criterion eliminates
 the larger values of $\Lambda_{\overline{\rm MS}}$ accepted by the criterion
 (\ref{crit}).

\begin{figure}
\begin{center}
\leavevmode
\mbox{\epsfig{file=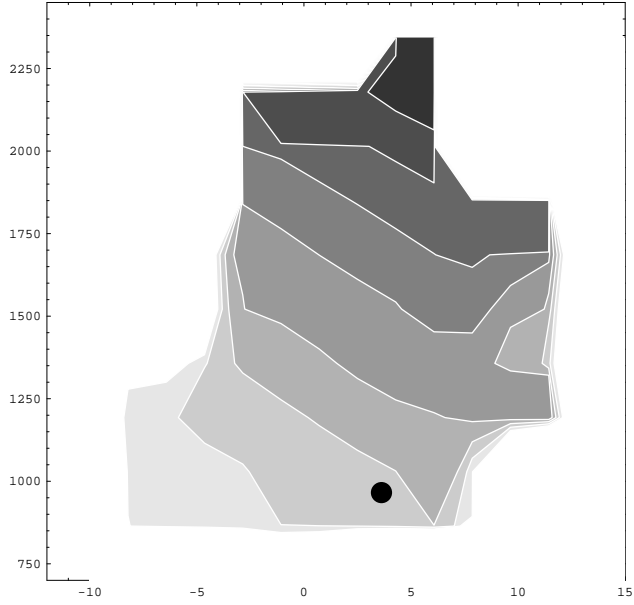,height=8cm}}
\end{center}
\caption{\it Domain of the ``good schemes'' in the $\widetilde\gamma_1, 
\widetilde\gamma_2$ plane. The level curves correspond to values of 
$\Lambda_{\overline{\rm MS}}$, ranging from $\simeq$ 380 MeV for the darkest
down to $\simeq$ 345 MeV. The black circle indicates the position of the 
$\widetilde {\rm MOM}$ scheme. The $\overline{\rm MS}$ scheme is far outside.}
\label{curv_niv}
\end{figure}

\begin{figure}[hbt]
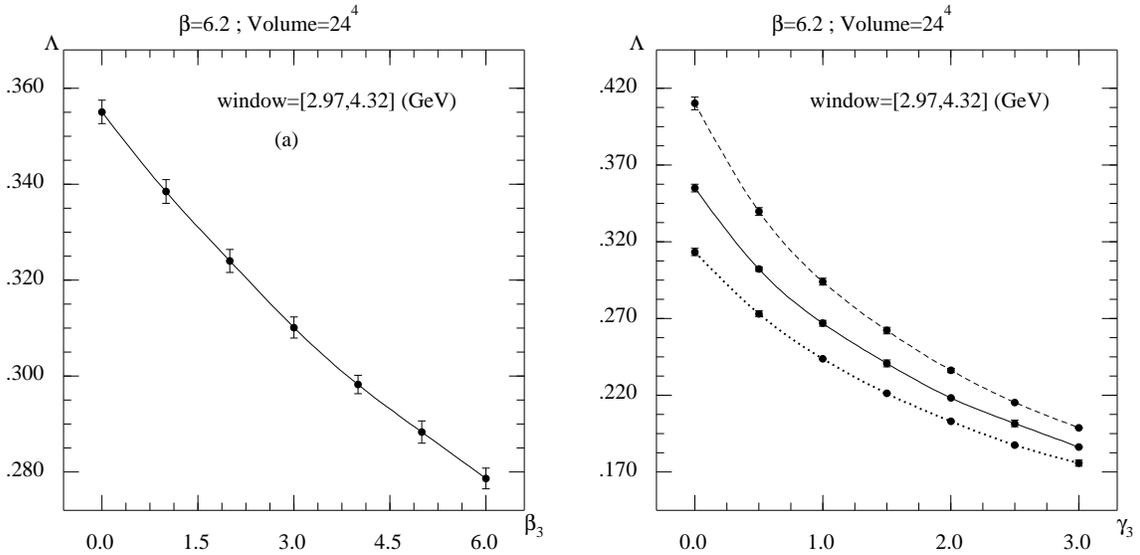

\begin{center}
\leavevmode
\mbox{\epsfig{file=lambda_beta3.eps,height=8cm}}
\mbox{\epsfig{file=lambda_gamma3.eps,height=8cm}}
\end{center}
\caption{\it Plot (a) shows $\Lambda_{\overline{\rm MS}}$ as a function of 
the unknown $\beta_3$ in the $\widetilde{\rm{MOM}}$ scheme
assuming $\gamma_3 = 0$. Plot (b) shows three different schemes as a function 
of $\gamma_3$ assuming $\beta_3=0$. The solid line if for the 
$\widetilde{\rm{MOM}}$ scheme. A tendency to converge is seen as
$\gamma_3$ increases }
\label{l_bg}
\end{figure}

The three conditions (\ref{crit}) and (\ref{crit2}) define
in the ($\widetilde\gamma_1,\widetilde\gamma_2$)-parameter space 
a domain of ``good schemes'' which is plotted in Fig. \ref{curv_niv}. The values of 
$\Lambda_{\overline{\rm MS}}$ vary between  $\simeq 345$ MeV and 
$\simeq 380$ MeV. We will take this as an indication about our systematic uncertainty.
However, this is not the whole story. All our analysis has shown that  the gluon
propagator {\sl is not asymptotic at three loops} in the present energy range.
The range $\simeq 345$ MeV to $\simeq 380$ MeV may be far from the real asymptotic 
value of $\Lambda_{\overline{\rm MS}}$ but it provides
  {\sl an effective three-loop $\Lambda_{\overline{\rm MS}}^{(3)}$ }. It is difficult 
  to estimate the distance of the latter from the former, ignorant as we are of
   the four-loops
  coefficients $\beta_3$ and $\gamma_3$. This distance can be large as we shall illustrate now.
We plot  in fig. \ref{l_bg}(a) the behaviour of $\Lambda_{\overline{\rm MS}}$,
 computed by obtaining the best fit over
the strong coupling constant in the $\widetilde{\rm{MOM}}$ scheme, as a function of the unknown
 $\beta_{3}$ assuming $\gamma_{3}=0$. As can be seen,
 a positive\footnote{The positive sign for $\beta_3$ is plausible
 as this is the sign of $\beta_i, i=0..2$ and as $\beta_{3\,\overline{\rm MS}}>0$.
 } $\beta_{3}$  would reduce the value of the four-loop effective 
$\Lambda_{\overline{\rm MS}}$ compared to the three-loop one. In fig. 
\ref{l_bg}(b) we plot assuming $\beta_3=0$ the behaviour of different schemes
among our ``good schemes'' as a function of $\gamma_3$: $\widetilde\gamma_3$
is computed from $\gamma_3$ and $\beta_3=0$ via eq.
(\ref{beta3}). We see that their
predictions for $\Lambda_{\overline{\rm MS}}$  tend to converge for positive
$\gamma_3$ where the value of $\Lambda_{\overline{\rm MS}}$ decreases. Of course
convergence between the schemes is an indication that we are closer to the
correct value of $\gamma_3$. 

Consequently, 
we {\sl expect that the asymptotic $\Lambda_{\overline{\rm MS}}$ will be 
smaller than the effective $\Lambda_{\overline{\rm MS}}^{(3)}$ which has been
estimated in this work}. 

\section{Conclusions}

Our initial aim has not been fulfilled. We cannot at the present 
energy scale give a reliable estimate of $\Lambda_{\overline{\rm MS}}$
from the gluon propagator because it has not yet reached its asymptotic regime
at three loops !
This is surprising since a scale as large as 5 GeV is often assumed, 
without further scrutiny, to be large enough to be asymptotic even
  at two loops ! The $\widetilde {\rm MOM}$ scheme is among our ``good schemes''. 
  It is amongst the schemes closest to asymptotic convergence. On the contrary 
  $\overline{\rm MS}$ is very far from convergence.  

We have learned an important lesson about the criteria of asymptoticity.
{\sl It is not enough to have a good fit over a large energy range with an asymptotic
formula to check asymptoticity}. Our fits are good because 
the error we introduce by ignoring higher orders is {\sl only logarithmic}.
Our energy range, as large as it may look,
corresponds to an increase of $\log q$ by only $\simeq 0.4$. Therefore
the higher loops effects can be mimicked by a simple rescaling of 
$\Lambda_{\overline{\rm MS}}$. The difference in the functional behavior 
of higher orders would
only be apparent on an energy range containing several ``e-foldings'', i.e. 
changes of $\log q$ by several units.  Such a study over several ``e-foldings''
up to a very high scale has been performed in \cite{luscher}.

  Combining our results for $\beta=6.2$ and $\beta=6.4$, we end up with 
  the following value for the
effective three-loops estimate:
\beq
\Lambda_{\overline{\rm MS}}^{(3)}= 353 \pm 2 ^{+25}_{-10}\, {\rm MeV}
\label{final}\eeq
  where the first error is statistical and the second is the systematic
  uncertainty estimated from the scheme dependence. 

From a study of the plausible effect of the fourth loop, we have argued 
that the real asymptotic $\Lambda_{\overline{\rm MS}}$ should lay below the
result in eq. (\ref{final}). 
   
We have discovered that the {\sl scheme independence of the result is a much more demanding 
criterion  of  asymptoticity} than the quality of fit. We have used many ``ad hoc'' schemes
simply defined by a couple of parameters $(\widetilde \gamma_1, \widetilde\gamma_2)$. 
This technique could be and
should be extended to other methods to compute $\alpha_s$ and more generally 
to any program which performs a matching to perturbative QCD, as long as several
renormalisation schemes can be used.

Finally we have proposed a new method
 to eliminate the hypercubic lattice artifacts,
namely to take the limit $p^{[4]}\to 0$, eq. (\ref{roiesnel}). We demonstrated that
this method is very efficient.
   
From the result of the present analysis we have undertaken a lattice calculation
at smaller lattice spacings, hoping to reach the asymptotic regime.

\section*{Acknowledgements.}

These calculations were performed on the QUADRICS QH1 located in the Centre de
Ressources Informatiques (Paris-sud, Orsay) and purchased thanks
to a funding from the Minist\`ere de l'Education Nationale and the CNRS.
D.B. acknowledges the Italian INFN, and J.R.Q. the Spanish Fundaci\'{o}n 
Ram\'{o}n Areces for financial support. We are indebted to Jacek Wosiek for
 useful discussions. We are specially indebted to Georges Grunberg for extensive
and illuminating discussions.

\end{document}